\documentclass{llncs}
\usepackage{makeidx}  % allows for indexgeneration
\usepackage{algorithm}
\usepackage{algpseudocode}
\usepackage{amsmath}
\usepackage{booktabs}
\usepackage{graphicx}
\usepackage{epsf}
\usepackage{amsfonts}
\usepackage{multirow}
\usepackage{bm}
\usepackage{subfigure}
\usepackage{url}
\usepackage{epsfig,graphics,psfrag,amssymb}
\begin{document}

\title{Optimization Matrix Factorization Recommendation Algorithm Based on Rating Centrality}
\author{Zhipeng Wu \and Hui Tian \and Xuzhen Zhu \and Shuo Wang}

\institute{State Key Laboratory of Networking and Switching Technology\\
Beijing University of Posts and Telecommunications, Beijing 100876, China\\\email{\{wzp,tianhui,zhuxuzhen,wangshuo16\}@bupt.edu.cn} }

\maketitle              % typeset the title of the contribution

\begin{abstract}
Matrix factorization (MF) is extensively used to mine the user preference from explicit ratings in recommender systems. However, the reliability of explicit ratings is not always consistent, because many factors may affect the user's final evaluation on an item, including commercial advertising and a friend's recommendation.
Therefore, mining the reliable ratings of user is critical to further improve the performance of the recommender system.
In this work, we analyze the deviation degree of each rating in overall rating distribution of user and item, and propose the notion of user-based rating centrality and item-based rating centrality, respectively.
Moreover, based on the rating centrality, we measure the reliability of each user rating and provide an optimized matrix factorization recommendation algorithm. Experimental results on two popular recommendation datasets reveal that
our method gets better performance compared with other matrix factorization recommendation algorithms, especially on sparse datasets.
\end{abstract}
\section{Introduction}
In recent years, the recommender system greatly promotes the development of e-commerce. A high-quality recommender system can help users quickly find what they like when they facing massive amounts of goods \cite{1a}, mitigate the problem of information overload \cite{2b}, and bring more economic benefits for sellers \cite{3c}. Consequently, in order to get more accurate recommendation results, many researchers have proposed various recommendation algorithms.

Among all the recommendation algorithms, collaborative filtering (CF) is a relatively simple and effective method to generate a list of recommendations for a target user \cite{4d}. One of the most popular methods of CF is user-based CF, it aims to find some users who have behavior records (such as commodity comment and browsing history) similar to the target user, and then recommend him those items that the similar users select \cite{5e}. Therefore, researchers have proposed a number of similarity calculation methods to find the similar users \cite{6f,7g,8h}, but the performance is barely satisfactory in the case of highly sparse data \cite{9a}.

Because of obtaining significant performance in the Netflix Prize competition, the model-based CF approaches gain remarkable development in recommender system due to their high accuracy and scalability \cite{9j,10h}, and matrix factorization model is the most representative one of them. It factorizes user-item rating matrix into two low rank matrices, the user-factor matrix and the item-factor matrix. Therefore, the original sparse rating-matrix can be filled by multiplication of two factor matrices. Inspired by the matrix factorization model in the competition, researchers have developed many improved algorithms successively: Funk \cite{11i} presented the regularized matrix factorization to solve the Netflix challenge and achieved a good result;
Sarwar \cite{12k} proposed incremental matrix factorization algorithm to make the recommender system highly scalable; Paterek \cite{13l} added user and item biases to matrix factorization for mining the interaction between user and item more accurately; Koren \cite{14m} integrated additional information sources in matrix factorization model, but the time complexity was very high;
Hu \cite{15n} proposed the concept of confidence level to measure user preferences in matrix factorization recommendation algorithm, which was only applicable to implicit feedbacks; He \cite{16o} pointed the missing data should be weighted by item popularity and provided the fast matrix factorization model; Meng \cite{18q} proposed weight-based matrix factorization and employed term frequency and inverse document frequency to find user's interests, but the method was only suitable for text data.

However, above methods fail to consider each explicit rating's reliability of user.
In general, users have their own tastes and opinions on an item. Although the explicit rating is made by user, not all ratings should be given the same weight \cite{9j}. For example, some users prefer to give items high scores, leading to their average scores much higher than the overall mean value. In contrast, other users only rate the favorite items and tend to give lower scores on other items \cite{21t}. In this situation, the preferences of the two types of users are distinctly different, if they give the same item same score, the reliability of the two scores should be carefully evaluated.

In our work, instead of using explicit ratings directly, we explore the reliability of each observed rating under limited user information. Firstly, we analyze the degree of deviation between each rating and the average score of user, propose the notion of user-based rating centrality. Similarly, according to the degree of deviation between each rating and the item average score, we define the item-based rating centrality. Then we combine two kinds of rating centrality to infer the reliability of a rating.
Furthermore, we provide an optimized matrix factorization algorithm based on the above analysis. Finally, we use stochastic gradient descent (SGD) learning algorithm to solve the optimization problem of the objective function. Several experiments are conducted on two classic recommendation datasets, and our method obtains better performance than other popular matrix factorization recommendation algorithms, indicating that it is feasible to mine the reliability of explicit ratings based on rating centrality.

The rest of the paper is organized as follows. Section 2 simply describes the matrix factorization recommendation algorithm. Our proposed approach which defines the rating centrality is introduced in detail in Section 3. In Section 4, we present the datasets and evaluation metrics in our experiments, and then analyze the experimental results. Finally, we draw the conclusion in Section 5.
\section{Preliminaries}
In this section, we first expatiate the problem discussed in this paper, and then give a brief introduction to traditional matrix factorization recommendation algorithms.
\subsection{Problem Definition}
In a general recommender system, we usually have $m$ users, $n$ items and the sparse user-item rating matrix $\textbf{R}\in \mathbb{R}^{m \times n}$. The each observed value $r_{u,i}$ of $\textbf{R}$ denotes the user $u$'s rating on item $i$, and $\hat{r}_{u,i}$ represents the predicted value of the user $u$ on item $i$. Given the interaction matrix $\textbf{R}$ of the user and item, the goal of the recommender system is to get the predicted values of items that the target user might interest.
\subsection{Matrix Factorization}
Matrix factorization algorithms have been extensively used to mine the interaction between user and item \cite{9j}.
Funk \cite{11i} points that the user-item rating matrix $\textbf{R}$ can be decomposed into two low rank matrices, the user-factor matrix and the item-factor matrix:
\begin{equation}
\begin{split}
\textbf{R}\approx \textbf{P}^{T}\textbf{Q}\ ,
\end{split}
\end{equation}
where $\textbf{P}\in \mathbb{R}^{k \times m}$ denotes user latent factor matrix, $\textbf{Q}\in \mathbb{R}^{k \times n}$ denotes item latent factor matrix and the parameter $k$ is the number of latent factors, in general, $k\ll \min(m,n)$.
Therefore, the predicted value $\hat{r}_{u,i}$ can be calculated as:
\begin{equation}
\hat{r}_{u,i}=\textbf{p}^{T}_{.,u}\textbf{q}_{\cdot,i}\ ,
\end{equation}
where $\textbf{p}_{.,u}$ is the $u$-th column of $\textbf{P}$ and $\textbf{q}_{.,i}$ is the $i$-th column of $\textbf{Q}$.
We can minimize the regularized squared error loss function $J$ to get latent factor matrices:
\begin{equation}
J=\sum_{(u,i)\in{T}}(r_{u,i}-\hat{r}_{u,i})^{2}
+\lambda (\left|\left| \textbf{p}_{.,u}\right|\right|^{2}_{F} +\left|\left| \textbf{q}_{.,i}\right|\right|^{2}_{F})\ ,
\end{equation}
where $\lambda$ is the parameter of regularization term that is to avoid over fitting, $\left|\left|\cdot \right|\right|_{F}$ denotes the Frobenius norm and $T$ is the training set of the (user, item) pairs. On the basis of this model, many improved matrix factorization algorithms have been proposed, for example, biased probabilistic matrix factorization \cite{13l}, weighted regularization matrix factorization \cite{20s}, coupled item-based matrix factorization \cite{20ss}, etc.

\section{Proposed Method}
In this section, we introduce the notion of rating centrality from the perspective of user and item respectively, which can be obtained easily even for sparse data. Based on the rating centrality, we present a strategy to compute the reliability of each rating and propose the optimized matrix factorization recommendation algorithm for further improving the accuracy of recommendation results.
\subsection{Notion}
The user-based rating centrality refers to the deviation degree between the user rating and the average score of user.
Even if two users have the same score on the same item, the user-based rating centrality of the two users may be totally different.
For example, user $a$ only rates whatever he likes, he will have a high average score, however,
another user $b$ tends to give items negative scores, consequently, the average score of the user $b$ is relatively low.
It is obvious that the preferences of the two users are completely different. In this case,
if user $a$ gives item $i$ a high score $r_{a,i}=5$ and user $b$ gives the same item same score $r_{b,i}=5$,
we can't regard the two ratings have the same reliability,
because user $a$ tends to give positive ratings and the average score of him is higher than user $b$.
Therefore, we define the user-based rating centrality $w_{u,i}^{U}$ to measure the reliability from user perspective:
\begin{equation}
w_{u,i}^{U}=\min(\frac{1}{\left| r_{u,i} -\bar{r}_{u} \right|}, r_{max})\ ,
\end{equation}
where $\bar{r}_{u}$ is the average score of user $u$ and $r_{max}$ is the maximum value of the rating scale. Because $r_{u,i}$ and $\bar{r}_{u}$ may be very close, to avoid the value of $w_{u,i}^{U}$ too large, we limit the max value of $w_{u,i}^{U}$ to $r_{max}$.

Moreover, the user-based rating centrality is just calculated from user perspective, if the quality of item $i$ is really good and user $b$ whose average score is low gives it a high rating, we should also suppose the rating has high reliability because the rating is consistence with item popularity. More exactly, if most users have preferences for the item $i$ and give it high scores, then the item $i$ will have a high average score. On the contrary, if the quality of item $j$ is poor, it will get plenty of negative feedbacks from the majority of users. Obviously, the characteristics of the two items are totally different.
In this case, if user $u$ gives item $i$ and $j$ same high score or same low score respectively, we should also consider the rating reliability from item perspective. Consequently, we define the item-based rating centrality $w_{u,i}^{I}$ to measure the deviation degree between the user rating and the average score of item:
\begin{equation}
w_{u,i}^{I}=\min(\frac{1}{\left| r_{u,i} -\bar{r}_{i} \right|}, r_{max})\ ,
\end{equation}
where $\bar{r}_{i}$ is the average value of the item $i$. Similarly, we limit the max value of $w_{u,i}^{I}$ to $r_{max}$.
In practical calculation, we can add a minimum value $\delta $ on the denominator in (4) and (5) to avoid denominator equals to zero, $\delta \ll 1$.

After obtaining the user-based and item-based rating centrality, we present a strategy to measure the reliability of a rating.
If both $w_{u,i}^{U}$ and $w_{u,i}^{I}$ are small values, that means the rating deviates from the overall distribution of user $u$ and item $i$, therefore, we suppose the rating has relatively low reliability.
However, if we get high values of $w_{u,i}^{U}$ and $w_{u,i}^{I}$, we will consider the rating reflects the real evaluation of user $u$ and item $i$, and give it a high weight. Hence, we can get the reliability of a rating $w_{u,i}$ from the following formula:
\begin{equation}
w_{u,i}=1+f(w_{u,i}^{U}w_{u,i}^{I})\ ,
\end{equation}
where $f(t)$ is a monotone increasing function that normalize the reliability. The bias $1$ is to avoid $w_{u,i}=0$ and maintain the data integrity. We will use three kinds of $f(t)$: $f(t)=tanh(t)$, $f(t)=sigmoid(t)$, $f(t)=t$, and conduct an experiment to compare the performance of them in Section 4.
\subsection{Prediction}
According to \cite{14m}, the prediction formula for the calculation of $\hat{r}_{u,i}$ in our method is defined as:
\begin{equation}
\hat{r}_{u,i}=b_{u}+b_{i}+\textbf{p}^{T}_{.,u}\textbf{q}_{\cdot,i}\ ,
\end{equation}
where $b_{u}$ is the bias of user $u$ and $b_{i}$ is the bias of item $i$. We consider that if a rating's reliability $w_{u,i}$ is low, then the influence of the rating should be reduced in training process. In other words, we pay more attention to the fitting of high reliability ratings in the process of optimizing the objective function. On this basis, we propose an optimized loss function with the weighted regularization which is to avoid the over fitting in the process of model training.
The adjusted regularized squared error loss function $J$ is as follows:
\begin{equation}
\begin{split}
J=\sum_{(u,i)\in{T}}(r_{u,i}-\hat{r}_{u,i})^{2}w_{u,i}
+\lambda (\bar{w}_{u}\left|\left| \textbf{p}_{.,u}\right|\right|^{2}_{F}+\bar{w}_{i}\left|\left| \textbf{q}_{.,i}\right|\right|^{2}_{F} + \bar{w}_{u}b_{u}^{2}+ \bar{w}_{i}b_{i}^{2}),
\end{split}
\end{equation}
where $\bar{w}_{u}$ and $\bar{w}_{i}$ denote the average reliability of user $u$'s ratings and item $i$'s ratings, respectively.

\subsection{Optimization}
In order to solve the problem of minimizing the loss function (8), we use SGD to learn the model parameters due to its high efficiency. First, for each observed rating $r_{u,i}$, we can get the the prediction error $e_{u,i}$:
\begin{equation}
\begin{split}
e_{u,i} = r_{u,i}-\hat{r}_{u,i}\ .
\end{split}
\end{equation}
Then we compute each parameter's partial derivative to get the direction of the gradient and next modify the parameters until the convergence is realized:
\begin{equation}
\begin{split}
b_{u}\leftarrow b_{u} -\eta(\lambda \bar{w}_{u}b_{u}-e_{u,i}w_{u,i}) \ ,
\end{split}
\end{equation}
\begin{equation}
\begin{split}
b_{i}\leftarrow b_{i} -\eta(\lambda \bar{w}_{i}b_{i}-e_{u,i}w_{u,i}) \ ,
\end{split}
\end{equation}
\begin{equation}
\begin{split}
\textbf{p}_{.,u}\leftarrow \textbf{p}_{.,u} -\eta(\lambda \bar{w}_{u}\textbf{p}_{.,u}-e_{u,i}w_{u,i}\textbf{q}_{.,i}) \ ,
\end{split}
\end{equation}
\begin{equation}
\begin{split}
\textbf{q}_{.,i}\leftarrow \textbf{q}_{.,i} -\eta(\lambda \bar{w}_{i}\textbf{q}_{.,i}-e_{u,i}w_{u,i}\textbf{p}_{.,u}) \ ,
\end{split}
\end{equation}
where $\eta$ is the learning rate. Our method is an improved matrix factorization algorithm based on rating centrality, so we call our method MFRC, and the specific algorithm flow is shown in Algorithm 1.
\begin{algorithm}[htp]
\caption{MFRC algorithm}%算法的名字
\hspace*{0.02in} {\bf Input: }%算法的输入， \hspace*{0.02in}用来控制位置，同时利用 \\ 进行换行
rating matrix \textbf{R}; reliability of each rating $w_{u,i}$; number of latent factors $k$;\\
\hspace*{0.02in} {\bf Output: }%算法的结果输出
latent factor matrix \textbf{P} and \textbf{Q}; biases $b_{u}$ and $b_{i}$; number of iterations $It$;
\begin{algorithmic}[1]
\State Randomly initialize \textbf{P} and \textbf{Q}; $b_{u}$ and $b_{i}$ are set to 0; $I = 0$; % \State 后写一般语句
\While{$I < It$} % While语句，需要和EndWhile对应
    \For{each observed rating $r_{u,i}$ in \textbf{R}} % For 语句，需要和EndFor对应
　　    \State update $b_{u}$ by equation (10);
        \State update $b_{i}$ by equation (11);
        \State update $\textbf{p}_{.,u}$ by equation (12);
        \State update $\textbf{q}_{.,i}$ by equation (13);
    \EndFor
    \State  $I = I + 1$;
\EndWhile
\end{algorithmic}
\end{algorithm}
\section{Experiments}
In this section, we introduce the datasets and evaluation metrics used in our experiments, and then analyze the experimental results in detail.
\subsection{Datasets}
MovieLens\footnote{https://grouplens.org/datasets/movielens/} dataset is one of the most prevalent datasets in recommender systems. In our experiments, we use two kinds of MovieLens datasets: MovieLens 100K and MovieLens 1M.
Each rating's range is from 1 to 5, and each user has rated at least 20 items. Table 1 shows the basic statistics of the two datasets.
\begin{table}
\caption{Statistics of Datasets}\label{tab1}
\begin{center}
\begin{tabular}{*{5}{c@{\;\;}}}
\hline\noalign{\smallskip}
 Dataset&Users&Items&Ratings&Sparsity\\
\noalign{\smallskip}
\hline
\noalign{\smallskip}
MovieLens 100K& 943&1,682  & 100,000&93.70\% \\
MovieLens\quad1M  & 6,040&3,952 & 1,000,209&95.80\% \\
\hline
\end{tabular}
\end{center}
\end{table}
\subsection{Benchmark Algorithms}
In our experiments, the benchmark algorithms contain probabilistic matrix factorization (PMF) \cite{11ii}, biased PMF (BPMF) \cite{14m}, alternating least squares with weighted regularization (ALSWR) \cite{20s} and AutoSVD \cite{27z}. They are closely
relevant to our work and achieve good results in recommender systems.
\subsection{Evaluation Metrics}
In order to evaluate the performance of the proposed method, we use root mean squared error (RMSE) and fraction of concordant pairs (FCP) to measure the accuracy of rating prediction.

RMSE is extensively used in measuring the accuracy of prediction, it is defined as:
\begin{equation}
\begin{split}
RMSE=\sqrt{\sum_{(u,i)\in{V}}(r_{u,i}-\hat{r}_{u,i})^{2}/|V|}\ ,
\end{split}
\end{equation}
where $V$ is the test set of (user, item) pairs and $|V|$ is the set size.

Another metric is FCP. Koren \cite{25x} supposed that the correct item ranking in recommender systems should also be considered.
In other words, if $r_{u,i} > r_{u,j}$ in test set, then this trend should be kept in prediction results.
Hence, FCP is defined as:
\begin{equation}
\begin{split}
FCP=\frac{\sum_{u}n_{c}^{u}}{\sum_{u}n_{c}^{u}+\sum_{u}n_{d}^{u}}\ ,
\end{split}
\end{equation}
where $n_{d}^{u}$ and $n_{d}^{u}$ denote the number of concordant pairs and discordant pairs of user $u$ respectively. Higher FCP means the more concordant pairs in test results. Therefore, we expect the recommendation algorithm has a high value of FCP when its RMSE is low.

\subsection{Results and Discussion}
\subsubsection{Impact of Normalization Function}
In this section, we compare the performance of three functions $f(t)$ in our model. We randomly choose 80\% of the original data as training set and the remaining as test set. The number of latent factors $k$ is from 20 to 100. From Fig.\,1, we can clearly see that ``$f(t)=tanh(t)$'' gets the best value of RMSE, and ``$f(t)=sigmoid(t)$'' performs sightly worse than ``$f(t)=tanh(t)$'', while ``$f(t)=t$'' performs the worst on both two datasets. This shows that the reliability of each rating should be normalized to a relatively small range. If we overemphasize on highly reliable ratings, we may lose much information from the remaining ratings and result in greater data sparsity. In terms of the performance of prediction, we will use ``$f(t)=tanh(t)$'' in the following experiments.
\begin{figure}[t]
    \begin{minipage}{0.5\linewidth}
        \centering
        \centerline{\epsfig{figure=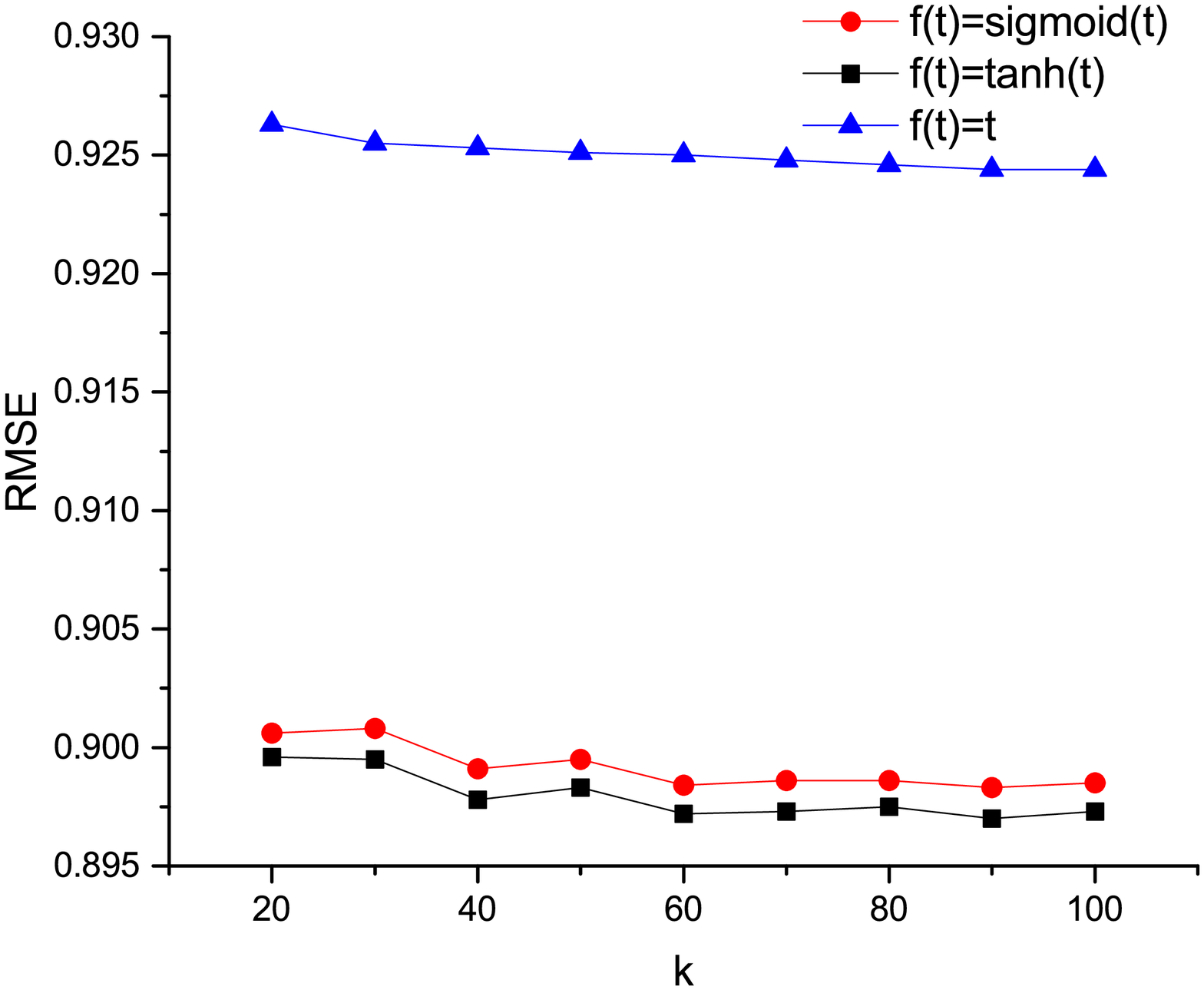,width=2.9in}}
        \centerline{(a) RMSE on MovieLens 100K}\medskip

    \end{minipage}%
    \begin{minipage}{0.5\linewidth}
        \centering
        \centerline{\epsfig{figure=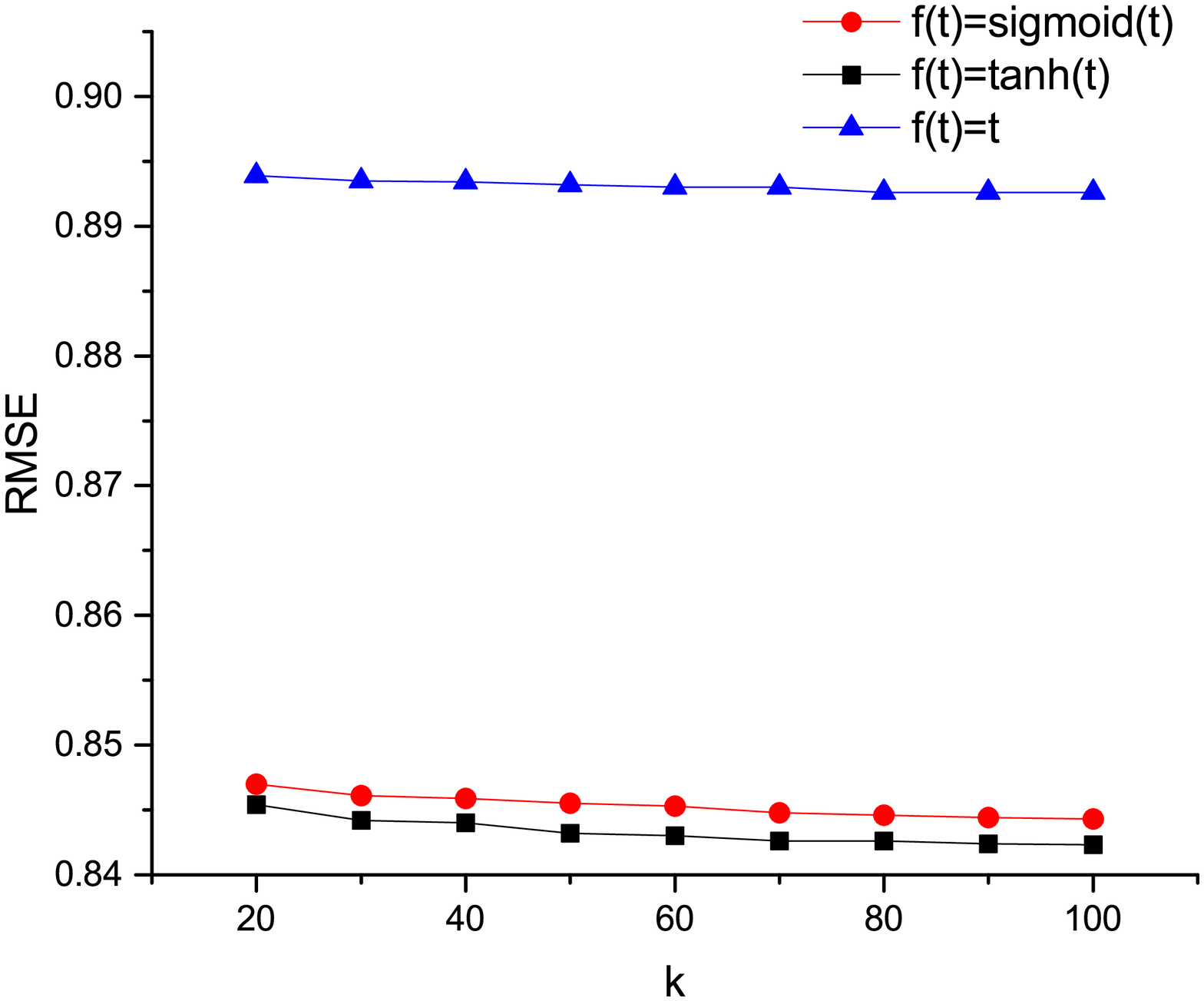,width=2.9in}}
        \centerline{(b) RMSE on MovieLens 1M}\medskip

    \end{minipage}
    \caption{Performance on two Datasets with different $f(t)$}
\end{figure}
\subsubsection{Impact of Number of Latent Factors}
In order to examine our method in depth, we compare our method with other benchmark algorithms under different number of latent factors $k$ ranging from 20 to 100. Similar to the previous section, we randomly choose 80\% of the original data as training set and conduct each experiment for five times, then calculate the average value of RMSE and FCP. In addition, the number of iterations is set to 100, $\lambda$ is set to 0.05 and $\eta$ is set to 0.005 on both two datasets.

Figure.\,2 shows that with the increase of $k$, the performances of all methods keep improving and eventually tend to be stable.
On MovieLens 100K dataset, from Fig.\,2(a), we can see that MFRC outperforms other benchmark algorithms on RMSE, which is at least 0.005 lower than BPMF and PMF. As for ALSWR, it is unstable with the change of $k$. In addition, Fig.\,2(b) shows that the FCP of all algorithms is up to 71\%, while MFRC has reached 74.5\% which is about 1\% higher than the second best algorithm. This indicates that our method not only has lower prediction error, but also has more correct ranked items pairs.
Similarly, on MovieLens 1M dataset which is more sparse than MovieLens 100K, the performance of our method is still the best one. Fig.\,1(c) shows the RMSE of MFRC maintains a gradual decline with the increase of k and at least 0.003 lower than that of BPMF. We can see from Fig.\,1(d) that the FCP of MFRC is significantly higher than that of other algorithms and is always maintained above 77.5\% when $k\geqslant 50$, while others are lower than 77.3\%.
According to above analysis, we can conclude that the performance of MFRC becomes better and gradually reaches a stable state with the increase of $k$, but obviously the computational complexity of matrix factorization is proportional to $k$. Therefore, we should consider the balance between accuracy and efficiency according to the actual situation.
\begin{figure}[t]
    \begin{minipage}{0.5\linewidth}
        \centering
        \centerline{\epsfig{figure=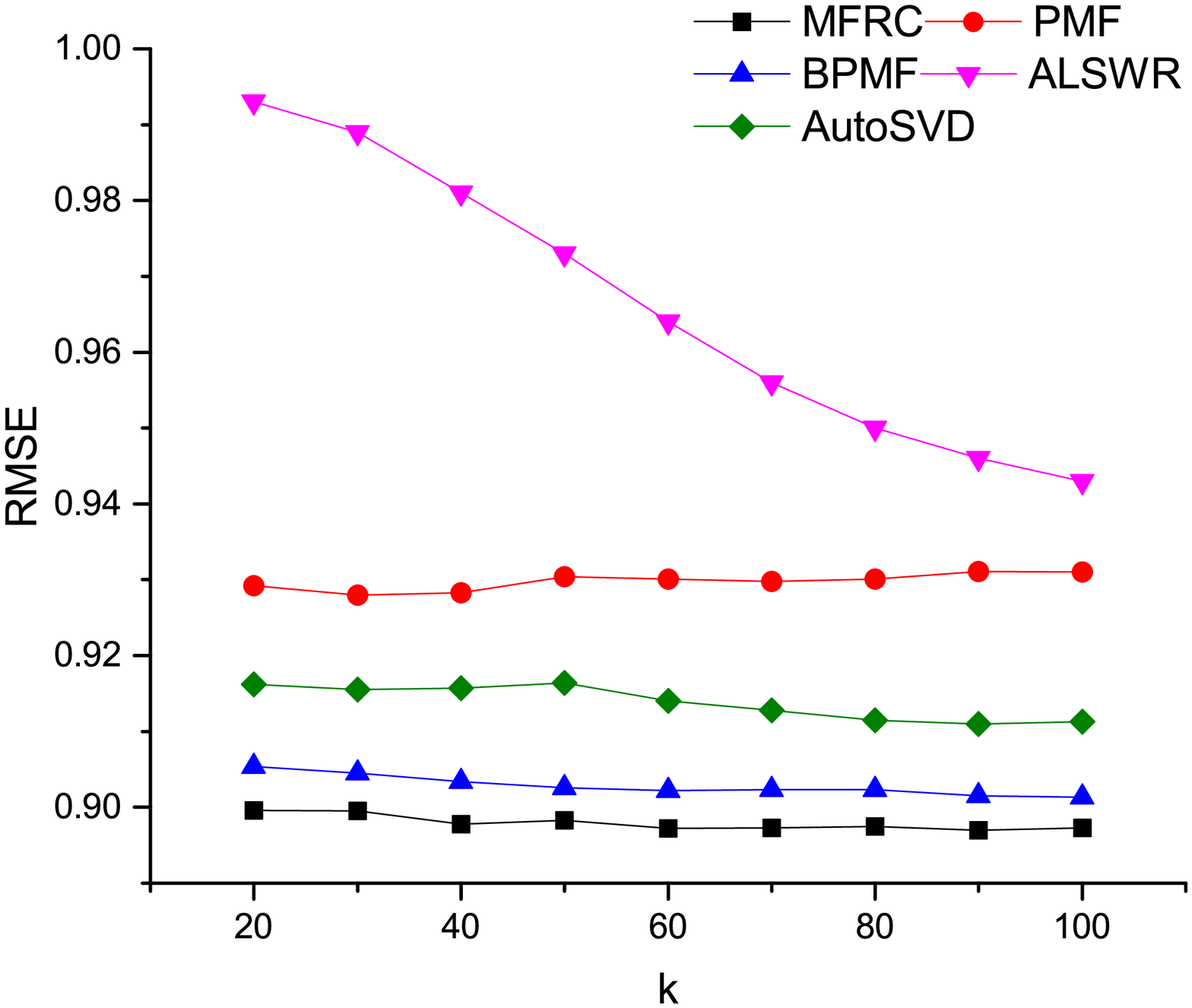,width=2.9in}}
        \centerline{(a) RMSE on MovieLens 100K}\medskip

    \end{minipage}%
    \begin{minipage}{0.5\linewidth}
        \centering
        \centerline{\epsfig{figure=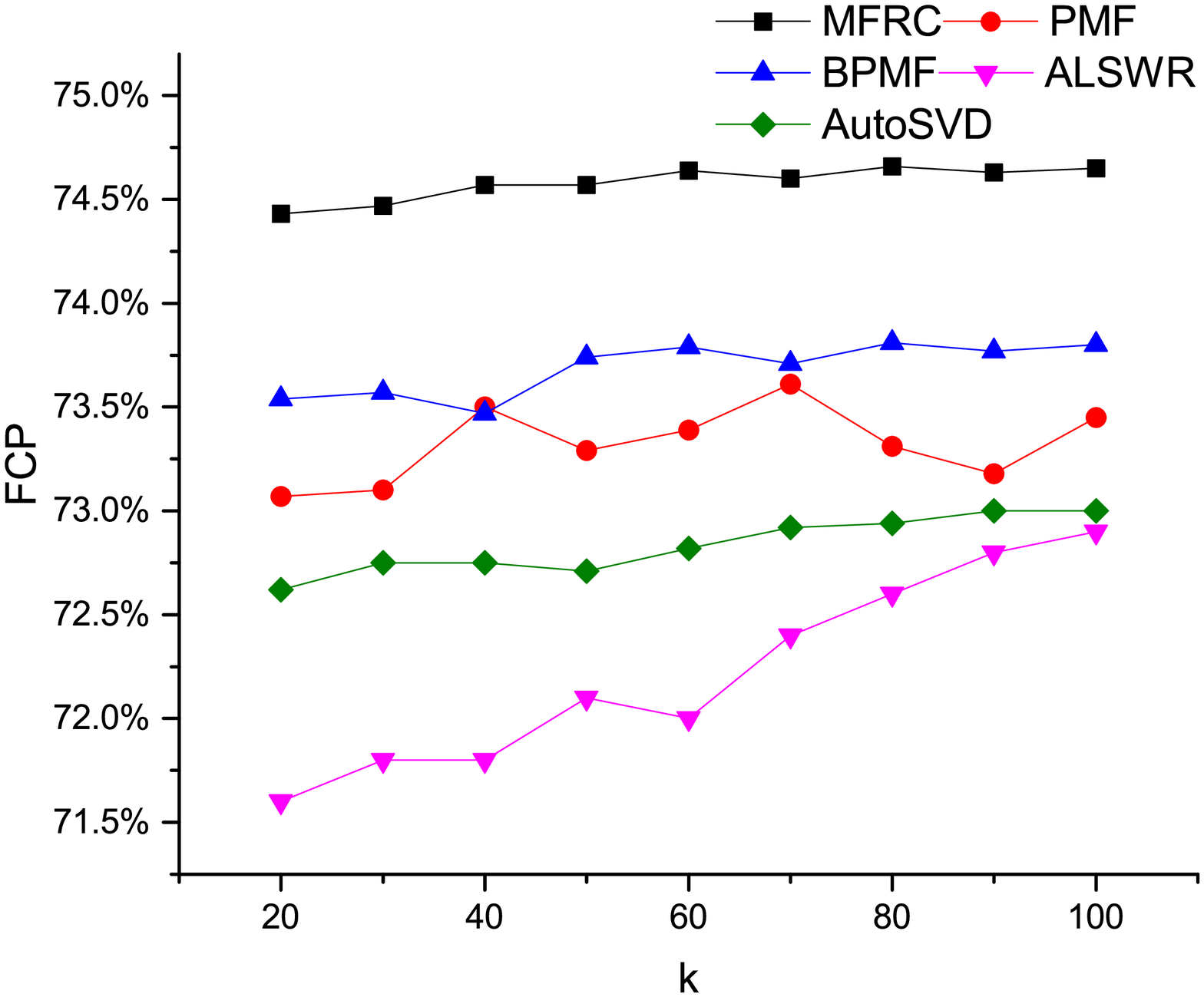,width=2.9in}}
        \centerline{(b) FCP on MovieLens 100K}\medskip

    \end{minipage}
    \begin{minipage}{0.5\linewidth}
        \centering
        \centerline{\epsfig{figure=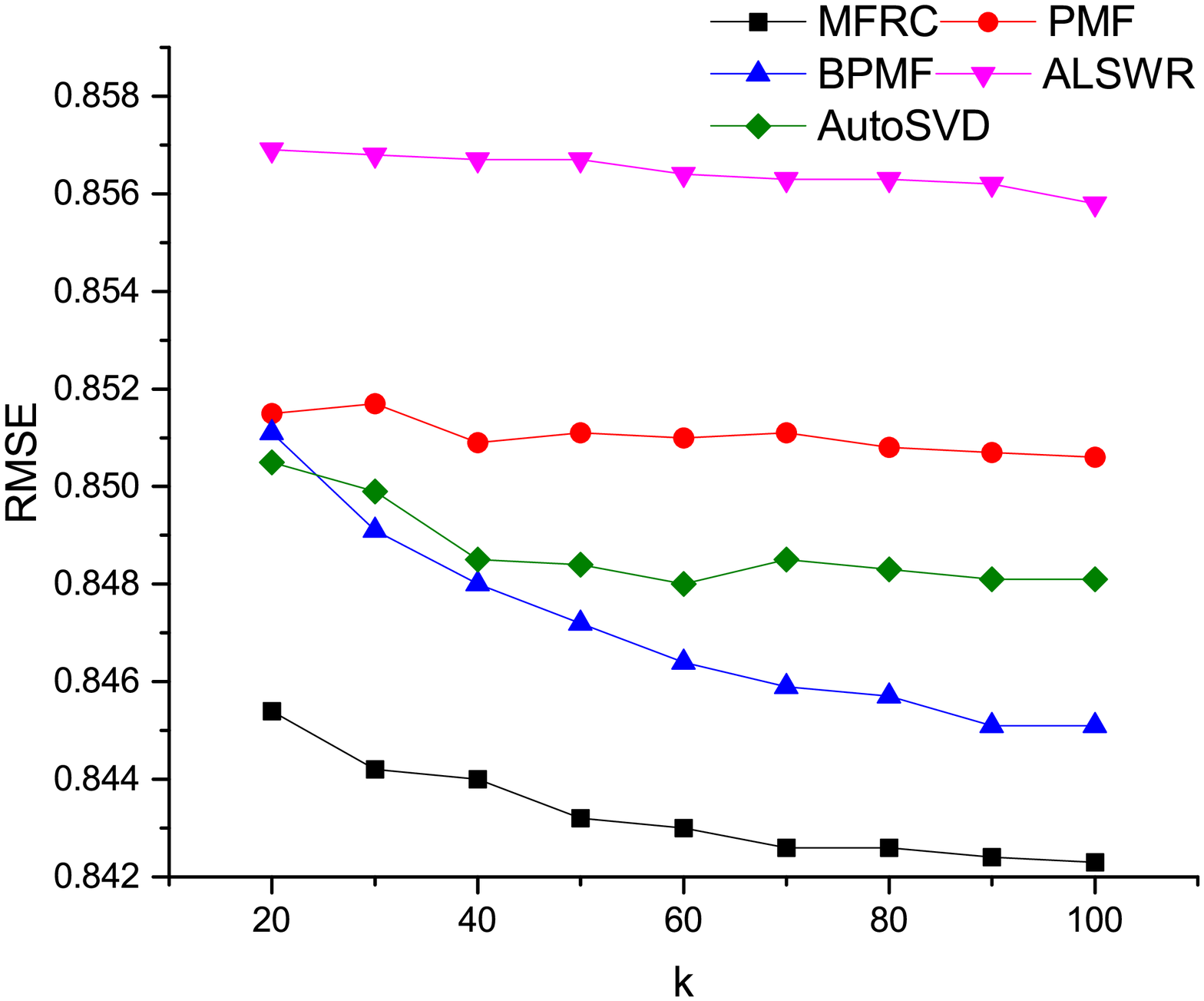,width=2.9in}}
        \centerline{(c) RMSE on MovieLens 1M}\medskip

    \end{minipage}%
    \begin{minipage}{0.5\linewidth}
        \centering
        \centerline{\epsfig{figure=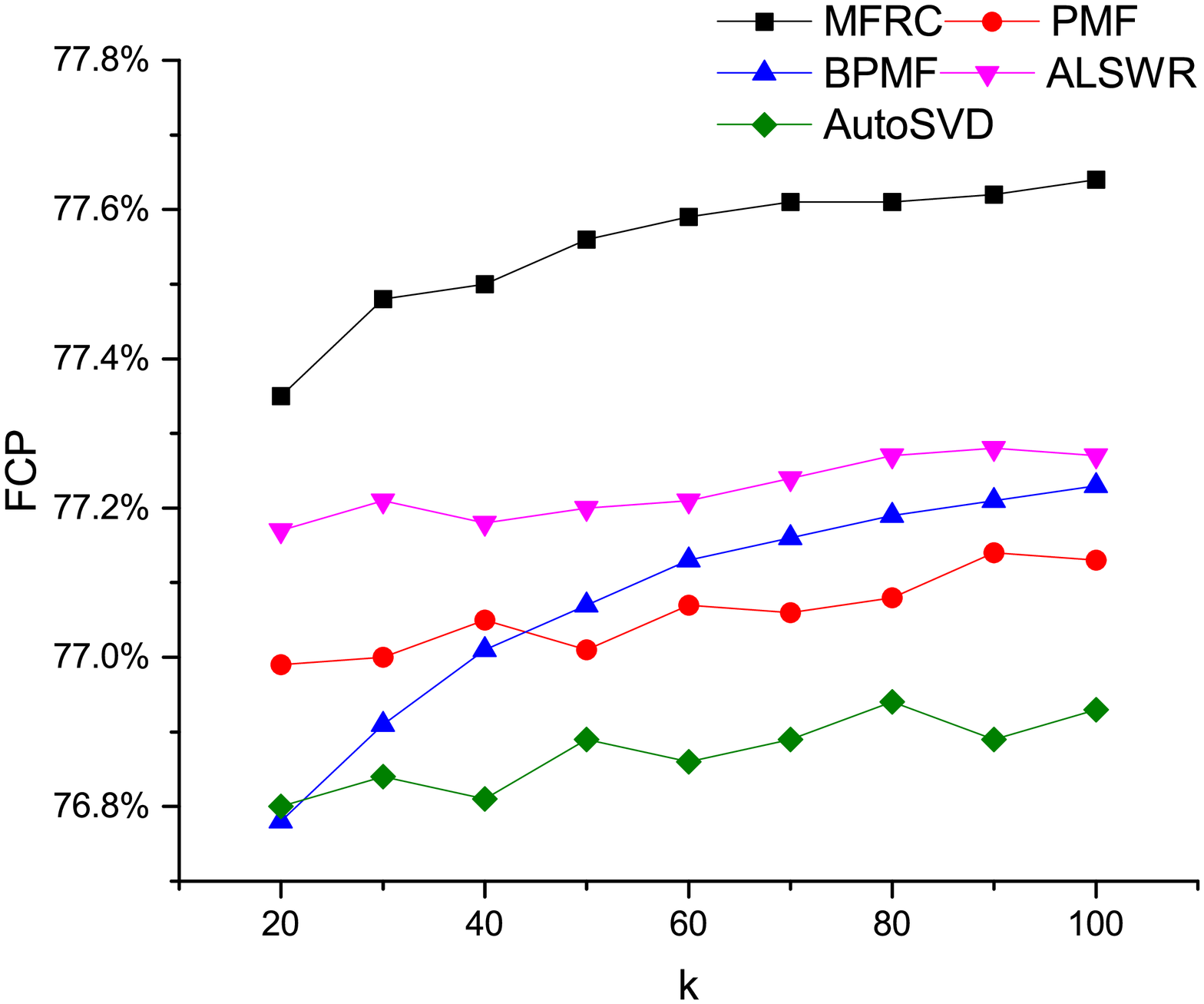,width=2.9in}}
         \centerline{(d) FCP on MovieLens 1M}\medskip

    \end{minipage}
    \caption{Performance on two Datasets under different latent factors $k$}
\end{figure}
\subsubsection{Impact of Sparsity}
Sparsity is one of the most important factors that affect the performance of the recommender system \cite{26y}.
To further evaluate our method, we change the proportion of training set. The training ratio $\alpha$ is set to 50\%, 60\%, 70\% and 80\%. The number of latent factors $k$ is set to 50.

Table 2 and Table 3 show the experimental results on two datasets, respectively. As we have expected, the sparsity of dataset greatly affects the performance of recommendation algorithm. Table 2 shows the results on MovieLens 100K dataset, we can see that from $\alpha=50\%$ to $\alpha=80\%$, the RMSE of MFRC is always maintained at a relatively low level and the FCP of MFRC increases steadily. Even though the improvement of performance becomes smaller and smaller, MFRC performs substantially well over all other benchmark algorithms. Through the comparison between BPMF and MFRC, we can find that it's effective to mine highly reliable ratings.
From Table 3, we can see clearly that on Movielens 1M dataset, our method outperforms significantly all methods discussed here under different data sparsity and the performance of traditional methods still has certain disparity compared to MFRC. When $\alpha=50\%$, the RMSE of MFRC is 0.008 lower than that of BPMF and 0.0221 lower than that of PMF. Similarly, the FCP of MFRC is kept at a high proportion with the increase of $\alpha$. In conclusion, our method that combined with rating centrality can make significantly less prediction error and get more concordant pairs on extremely sparse datasets.
\begin{table}
\caption{Performance on MovieLens 100K under different training ratio $\alpha$}
  \begin{center}

  \label{tab:performance_comparison}
    \begin{tabular}{*{7}{c@{\quad}}}
    \midrule
    Metric&$\alpha$ &PMF&BPMF&ALSWR&AutoSVD&MFRC\\
    \midrule
    \multirow{4}{*}{RMSE}
                    &50\%&0.9751& 	0.9358& 	1.0112& 	0.9417& 	\bf0.9270\cr
                    &60\%&0.9557& 	0.9196& 	1.0063& 	0.9317& 	\bf0.9163\cr
                    &70\%&0.9412& 	0.9116& 	0.9891& 	0.9245& 	\bf0.9067\cr
                    &80\%&0.9304& 	0.9026& 	0.9730& 	0.9164& 	\bf0.8983\cr

    \midrule
    \multirow{4}{*}{FCP}

                    &50\%&71.35\%& 	71.61\%& 	70.11\%& 	70.51\%& 	\bf72.53\%\cr
                    &60\%&72.03\%& 	72.37\%& 	70.23\%& 	71.24\%& 	\bf73.20\%\cr
                    &70\%&72.60\%& 	72.90\%& 	71.06\%& 	71.77\%& 	\bf73.74\%\cr
                    &80\%&73.29\%& 	73.74\%& 	71.60\%& 	72.71\%& 	\bf74.57\%\cr

    \midrule

    \end{tabular}
    \end{center}
\end{table}

\begin{table}
\caption{Performance on MovieLens 1M under different training ratio $\alpha$ }
  \begin{center}

  \label{tab:performance_comparison}
    \begin{tabular}{*{7}{c@{\quad}}}
    \midrule
    Metric&$\alpha$ &PMF&BPMF&ALSWR&AutoSVD&MFRC\cr
    \midrule
    \multirow{4}{*}{RMSE}
                    &  50\%&0.8830 	&0.8689 	&0.8802	    &0.8753& 	\bf0.8609\cr
                    &  60\%&0.8698 	&0.8597 	&0.8681 	&0.8653& 	\bf0.8531\cr
                    &  70\%&0.8584 	&0.8514 	&0.8593 	&0.8560& 	\bf0.8465\cr
                    &  80\%&0.8511 	&0.8472 	&0.8567 	&0.8484&    \bf0.8432\cr

    \midrule
 \multirow{4}{*}{FCP}
                    &  50\% &75.33\%& 	75.90\% &	75.75\% &   75.38\% 	&\bf76.67\%\cr
                    &  60\% &76.05\%& 	76.41\% &	76.47\% &	75.99\% 	&\bf77.08\%\cr
                    &  70\% &76.66\%& 	76.86\% &	76.97\% &	76.53\% 	&\bf77.44\%\cr
                    &  80\% &77.01\%&	77.07\% &	77.20\% &	76.89\%     &\bf77.56\%\cr

    \midrule

    \end{tabular}
    \end{center}
\end{table}
\section{Conclusion}
In this work, for getting more accurate recommendation results, we mine the reliable ratings of user from limited data, and propose an optimized matrix factorization recommendation algorithm based on rating centrality of user and item. Different from traditional matrix factorization recommendation algorithms which fail to consider the reliability of each user rating, in our method, we define the notion of user-based rating centrality and item-based rating centrality, and then combine them to measure the reliability of each rating. On this basis, we introduce the reliability into traditional matrix factorization objective function and make an optimized adjustment. Our extensive experimental results demonstrate that MFRC obtains less prediction error and more concordant pairs compared with other popular matrix factorization recommendation algorithms, especially on highly sparse datasets. We can conclude that our method based on rating centrality can find the reliable rating from user's explicit ratings and get significant performances in recommender systems.
\section*{Acknowledgement}
This work was supported by the National Natural Science Foundation of China (No.61602048) and the Fundamental Research Funds for the Central Universities(No.NST20170206).

%
% ---- Bibliography ----
%
\bibliography{my}
\bibliographystyle{splncs03}

\end{document}